\documentclass[prl,preprint,superscriptaddress]{revtex4}

\usepackage{graphicx}
\usepackage{dcolumn}
\usepackage{bm}

\usepackage{color}

\begin{document}


\title{Temperature-driven modification of surface electronic structure on bismuth, a topological border material}

\author{Y. Ohtsubo}
\email{y_oh@fbs.osaka-u.ac.jp}
\affiliation{Graduate School of Frontier Biosciences, Osaka University, Suita 565-0871, Japan}
\affiliation{Department of Physics, Graduate School of Science, Osaka University, Toyonaka 560-0043, Japan}

\author{Y. Yamashita}
\author{J. Kishi}
\affiliation{Department of Physics, Graduate School of Science, Osaka University, Toyonaka 560-0043, Japan}

\author{S. Ideta}
\author{K. Tanaka}
\author{H. Yamane}
\affiliation{Institute for Molecular Science, Okazaki 444-8585, Japan}

\author{J. E. Rault}
\author{P. Le F\`evre}
\author{F. Bertran}
\affiliation{Synchrotron SOLEIL, Saint-Aubin-BP 48, F-91192 Gif sur Yvette, France}

\author{S. Kimura}
\email{kimura@fbs.osaka-u.ac.jp}
\affiliation{Graduate School of Frontier Biosciences, Osaka University, Suita 565-0871, Japan}
\affiliation{Department of Physics, Graduate School of Science, Osaka University, Toyonaka 560-0043, Japan}

\date{\today}

\begin{abstract}
Single crystalline bismuth (Bi) is known to have a peculiar electronic structure which is very close to the topological phase transition.
The modification of the surface states of Bi depending on the temperature are revealed by angle-resolved photoelectron spectroscopy (ARPES).
At low temperature, the upper branch of the surface state merged to the projected bulk conduction bands around the $\bar{M}$ point of the surface Brillouin zone (SBZ).
In contrast, the same branch merged to the projected bulk valence bands at high temperature (400 K).
Such behavior could be interpreted as a topological phase transition driven by the temperature, which might be applicable for future spin-thermoelectric devices.
We discuss the possible mechanisms to cause such transition, such as the thermal lattice distortion and electron-phonon coupling. 
\end{abstract}

\maketitle

\section{Introduction}

After the discovery of the three-dimensional topological insulator (TI) \cite{Hsieh08}, the electronic structure of such topological materials has been studied very extensively in these days as a promising template for future spintronic technologies \cite{Hasan10, Manchon15, Han18}.
The classification of the materials to topological or normal ones is based on the symmetry operations of bulk electronic structure.
For example, the parity eigenvalues of time-reversal operation determine the $Z_2$ topological order (TO) of insulators and semimetals with finite bandgap at any $k$ points \cite{Fu07, Teo08}.
Protected by such TO, the topological surface states dispersing between the  bulk valence bands (BVB) and bulk conduction bands (BCB) always appear irrespective to the detailed atomic structure of the surface.

The topological character of the surface electronic states are governed by the TO of bulk electronic structure.
Therefore, qualitative modifications of the bulk electronic structure as well as its TO is reflected to the surface states. 
Such topological phase transitions have been reported by applying the magnetic order \cite{Xu12, Chang13, Hirahara17}, chemical substitution \cite{Xu11, Sato11}, or changing the symmetry group itself \cite{Wojek15}.
In recent days, topological phase transitions are gathering much interest, because various new topological phenomena are expected to appear during such transition, such as anomalous quantum Hall effect emerging with magnetic topological phase transition \cite{Chang13, Hirahara17}.

In this work, we tried to trace the possible topological phase transition of single-crystalline Bi driven by temperature.
Single crystal Bismuth (Bi) is known to have a peculiar electronic structure that is very close to the topological phase transition \cite{Hsieh08, Teo08, Hirahara12, Ohtsubo13, Benia15, Aguilera15, Ito16, Ohtsubo16}.
The modification of the surface electronic states of Bi depending on the temperature are revealed by angle-resolved photoelectron spectroscopy (ARPES).
At low temperature, the upper branch of the surface electronic state merged to the projected BCB around the $\bar{M}$ point of the surface Brillouin zone (SBZ).
In contrast, the same branch merged to the projected BVB at high temperature (400 K).
Such behaviour could be interpreted as a topological phase transition, as explained in the following section.
The possible mechanisms to cause the transition, such as the thermal lattice distortion \cite{Hirahara12, Aguilera15, Ohtsubo16} and electron-phonon coupling \cite{Garate13} are discussed. 
Such new mechanism to undergo the topological phase transition might be useful for future spin-dependent thermoelectric devices.

\section{Topological order of single-crystal Bismuth}

Single crystal of Bi has a rhombohedral unit cell, forming the bilayered honeycombs stacking along [111] (see Fig. 1 (a)).
The bulk electronic structure of Bi is a typical semimetal with finite bandgap at any $k$ points and small hole and electron pockets at $T$ and $L$ of the Brillouin zone shown in Fig. 1 (b), respectively \cite{Liu95}.
It is well known that Bi becomes semiconductor by alloying with small amount of Sb and such alloy is the first three-dimensional TI discovered by ARPES \cite{Hsieh08}, as theoretically predicted \cite{Fu07, Teo08}.
Figure 1 (c) depicts the qualitative dispersion of the BiSb alloy; the upper branch of the surface states connects the BVB and BCB continuously.
Such dispersion of topological surface states were verified by ARPES \cite{Hsieh08, Benia15}.

In contrast, TO of the single-crystal Bi is still controversial.
From various theoretical models \cite{Fu07, Teo08, Liu95, Aguilera15}, it is calculated to be a normal semimetal.
Based on the normal TO, both branches of the surface states should merge to the same projected bulk bands (the case depicted in Fig. 1 (d)) or degenerate with each other at both time-reversal-invariant momenta ($\bar{\Gamma}$ and $\bar{M}$ in this case).
However, the surface-state dispersion of pure Bi observed by ARPES showed the continuous dispersion of a surface branch between BVB and BCB, indicating the same feature as topological BiSb \cite{Ohtsubo13, Ito16}.

Although the TO of single-crystal Bi is not clear as explained above, it is commonly accepted that Bi is very close to the topological phase transition.
The transition occurs by the bulk band inversion at $L$ (see Table I) and the bandgap there is only $\sim$15 meV.
Therefore, very small modification could change the TO of Bi.
It would also be the main reason why the determination of the TO of pure Bi is still controversial.
Based on the empirical tight-binding calculation, the lattice distortion within 2 \% causes such transition, as shown in Fig. 2 \cite{Ohtsubo16}.
In parallel, the state of the art first-principles calculation also predicted that 0.4 \% distortion is enough \cite{Aguilera15}.
Actually, the surface electronic structure of Bi with tensile strain has been already reported to exhibit topological-semimetal character \cite{Hirahara12}.
However, to the best of our knowledge, surface states of Bi indicating normal TO, caused by the bulk band inversion at $L$, has never been observed experimentally.

\section{Experimental Methods}
The clean Bi(111) surfaces were prepared by repeated cycles of argon ion sputtering at 0.5 keV and annealing up to 450$\pm$20 K until a sharp low-energy electron diffraction (LEED) pattern was observed as shown in Fig. 3 (a).
In this work, a multichannel-plate-amplified LEED equipment was used.
The in-plane surface lattice distortion was checked by the LEED pattern at different temperatures.
The sample temperatures for LEED and ARPES measurements were monitored by a diode temperature sensor attached close to the sample.
ARPES measurements were performed with a He lamp and synchrotron radiations at the CASSIOP\'EE beamline of synchrotron SOLEIL (photon energies ranged from 25 to 80 eV).
The overall energy resolutions were 10 meV with the He lamp and 15 meV for synchrotron radiation, evaluated by the width of the Fermi edge of Mo foils attached to the sample.

\section{Results and Discussion}

\subsection{Thermal expansion of surface lattice constant}
The thermal expansion of the surface lattice constant was checked by the LEED patterns at 30 and 400 K, as shown in Fig. 3.
For these LEED patterns, the sample was fixed at the same position in front of LEED and its temperature was tuned there.
At both temperatures, sharp and bright electron diffraction spots indicating three-fold symmetry of the Bi(111) surface were clearly observed.
For the quantitative analysis of the thermal expansion, the line profiles of the LEED pattern were obtained as shown in Figs. 3 (c-e).
The intensity vanishes at the centre of the profiles (670-730 pixels) because this area is shaded by an electron gun.
While the thermal background is higher at 400 K, the sharp spots were found in both profiles.
The peak positions of each spot were obtained by fitting the profile with a Gaussian function and linear background as shown in Figs. 3 (d, e).
From this data, we found that the in-plane lattice expansion is $\sim$0.2 \% from 30 to 400 K.
This value is in the same order as the bulk thermal expansion obtained by x-ray diffraction \cite{Fischer78}, 0.4 \%, and also in the same order as the required lattice distortion for topological phase transition based on the first-principles calculation \cite{Aguilera15}.


\subsection{Surface electronic structure modification caused by temperature}
Figure 4 shows the ARPES band dispersions along $\bar{\Gamma}$-$\bar{\rm M}$ at 20 K and 400 K.
At low temperature (Fig. 4 (a)), the two branches of the surface bands, $S1$ and $S2$ forming an electron pocket at $\bar{\Gamma}$ and a shallow hole pocket around 0.3 \AA$^{-1}$, respectively, were observed, consistent with the earlier results \cite{Ohtsubo13, Ito16, Hirahara08}.
It has been already reported based on spin-resolved ARPES that $S1$ and $S2$ are the surface spin-split branches with the spin polarizations towards the opposite orientations to each other \cite{Hirahara08}.
Both $S1$ and $S2$ merges into BVB at $\bar{\Gamma}$, consistent with what is depicted in Figs. 1 (c, d).
At high temperature (400 K, Fig. 4 (b)), the qualitative behaviour of the $S1$ and $S2$ are the same as those at low temperature.
However, $S2$ around $\bar{\Gamma}$ moves to lower binding energies, as shown in the EDCs at 0.05 \AA$^{-1}$ in Fig. 4 (c): both $S1$ and $S2$ peaks are visible at 20 K but they merges to a broad single peak at 400 K because of the upward shift of $S2$.
It should be noted that such band shift is not rigid.
Away from $\bar{\Gamma}$, the $S2$ band moves oppositely, towards the higher binding energies, as shown in the EDC at 0.36 \AA$^{-1}$ in Fig. 4 (c) and the smaller values of the second $k_{\rm F}$, $\sim$ 0.3 \AA$^{-1}$, in Fig. 4 (d).
Such non-rigid deformation of the surface bands might be due to the temperature-dependent small change of the bilayer-buckling factor $\mu$ in the Bi crystal structure \cite{Fischer78}.

Figure 5 is the ARPES band dispersions at low (20--30 K) and high (400 K) temperatures, measured around $\bar{\rm M}$.
At low temperature shown in Fig. 5 (a), the upper branch, $S1$ appears again below the Fermi level and shows nearly flat dispersion at $\sim$ 20 meV.
It looses the photoelectron intensity at $\bar{\rm M}$, suggesting its merging into projected bulk bands.
The same behaviour occurs for $S2$ around $-$0.2 \AA$^{-1}$ from $\bar{\rm M}$.
On the clean surface of Bi(111), it was reported that the energy positions of BVB and BCB are nearly the same as the bulk ones; $\sim$25 meV below the Fermi level for the bottom of BCB and $\sim$40 meV for the top of BVB \cite{Ohtsubo13}.
According to this, $S1$ does not couple to BVB but to BCB at $\bar{\rm M}$.
Such behaviour, $S1$ connecting the BVB at $\bar{\Gamma}$ and BCB at $\bar{\rm M}$, agrees with the topological case depicted in Fig. 1 (c) and is consistent with the earlier experimental results \cite{Ohtsubo13, Ito16}.
In addition to the surface bands, an edge of the broad photoelectron intensities is observed around $\bar{\rm M}$, as guided by the dotted lines in Figs. 5 (a, b).
It would be the upper edge of the projected BVB.
Actually, the top of the edge in Fig. 5 (a) is around 50 meV, close to the expected position of the top of BVB (40 meV).
$S1$ is clearly separated from such edge at low temperature.

At higher temperature, as shown in Fig. 5 (b), the dispersion of $S1$ changes qualitatively.
In contrast to the nearly flat dispersion at low temperature, $S1$ at 400 K disperses nearly linearly from $k_{\rm F}$ ($\sim$-0.25 \AA$^{-1}$) to $\bar{\rm M}$, reaching the binding energy around 60 meV.
The top of the BVB expected from the edge of broad photoelectron intensity also moved slightly downward and apparently $S1$ merges into there.
The change of the surface-band dispersion was also observed along $\bar{\rm M}$-$\bar{\rm K}$ shown in Figs. 5 (c, d).
The EDC peak corresponding to $S1$ is at $\sim$20 meV at 20 K and moves downwards to $\sim$60 meV at 400 K.
At 400 K, the $S1$ band also appears to be merged into the valence bands, as shown in Fig. 5 (d).
Such dispersion of $S1$, merging into BVB both at $\bar{\Gamma}$ and $\bar{\rm M}$, agrees with what is expected for the topologically normal case as depicted in Fig. 1 (d).
Therefore, it is suggested that the topological phase transition from topological to normal phase occurs depending on the temperature.

\subsection{Possible topological phase transition}
In order to pursue the temperature dependent change of the surface band $S1$, we traced the ARPES peak positions at various temperatures as shown in Fig. 6.
Near $k_{\rm F}$, the peak positions corresponding to $S1$ stay at nearly the same binding energies with elevating temperatures as shown in Fig. 6 (a).
However, close to $\bar{\rm M}$, the shift becomes evident, as shown in the rest of Fig. 6.
The shift starts at around 100 K and is monotonic up to 400 K.
The energy shift of $S1$ from 30 to 400 K is $\sim$30 meV.
This value is twice larger than the bulk bandgap at $L$ (corresponding to $\bar{\rm M}$ in surface Brillouin zone).
Therefore, such energy shift would be enough to suppose the bulk bandgap inversion at $L$ to cause the topological phase transition.
If such topological phase transition actually occurred, the bulk band gap should be closed at a critical temperature and open again above there.
However, from the current data, we could not find any critical temperature around where the surface bands behaves differently.
ARPES cannot observe the bulk bands of Bi in detail with the current experimental condition, because the bright surface bands are always observed at the same time.
Therefore, it is difficult to find out the specific temperature for the supposed topological phse transition.

Here we'd like to discuss the possible origin of the topological phase transition depending on the temperature.
The first possibility is the thermal lattice distortion suggested by theoretical models \cite{Aguilera15, Ohtsubo16}.
The surface lattice expansion was evaluated as $\sim$0.2 \% by LEED.
Although this value is one order of magnitude smaller than the expected value from a tight-binding model \cite{Ohtsubo16}, it is at the same order as what is expected from first-principles calculation \cite{Aguilera15}.
However, the distortion direction is the opposite; the first principles model expected the transition with tensile strain.
Moreover, the first principles model predicts the surface electronic states with normal TO for Bi at low temperature, in contrast to the ARPES experimental results.
Such discrepancies might be reconciled by assuming additional correction factor missing in the current first-principles model, which inverts the TO of Bi to topological.
In such case, the required lattice distortion might also be inverted to be the expansion.
However, we have to admit that this scenario requires the unknown and arbitrary ``correction'' to the current theoretical models and that we do not have any explicit origin of such factor.

Alternatively, band inversion and topological phase transition assisted by phonon excitation is also theoretically expected \cite{Garate13}.
In this model, thermally excited phonon is coupled to bulk electronic states and renormalizes the size of the bandgap.
Such electron-phonon coupling could reduce the size of the bandgap and could even close and invert the gap, if the size of the bandgap is small enough.
Apparently, this model agrees with the current case, Bi with very small bulk band gap of $\sim$15 meV.
However, in order to verify this electron-phonon coupling model, further experiments to trace the phonon excitations depending on the temperature and the bulk bandgap inversion itself are required.

\subsection{Comparison with known bulk electronic properties}

The bulk electronic properties of single crystal Bi depending on the temperature have been already studied in early days by magnetoreflection analysis \cite{Vecchi74, Vecchi74-2}.
In such works, monotonous expansion of the bulk bandgap at $L$ without closing were reported.
At first glance, it contradicts to the current ARPES results suggesting the topological phase transition.
However, in these works, the rigid two-band model was used and all the change of the experimental data depending on the temperature were explained as a change of bulk bandgap.
Such model is not consistent with the non-rigid band shift that we observed in Figs. 4 and 5.
Therefore, in order to understand the thermally driven modification and possible topological phase transition of Bi, further study is desirable, especially to explain the non-rigid band modification depending on the temperatures higher than room temperature.


Although more studies are required to verify it, the thermally-driven topological phase transition we propose is quite attractive for future spin-dependent thermoelectric technologies, because it means thermal gradient could make the interface between the topological and normal insulators.
Similar to the surface of the topological insulator, such interface should also hold the topological interface states passing the spin current.
Therefore, it would be another mechanism to convert the thermal gradient to spin current, parallel to the spin Seebeck effect \cite{Uchida08}.
The relationship between thermal topological interface and spin Seebeck effect would be an analogy of that between Rashba-Edelstein effect and spin-Hall effect \cite{Han18}.


\section{Summary}
The temperature-driven modification of the surface states of Bi, which is known to be very close to the topological phase transition, is studied by ARPES.
At low temperature (20--30 K), the upper branch of the surface state merged to the projected bulk conduction bands around the $\bar{M}$ point of the surface Brillouin zone (SBZ).
In contrast, the same branch merged to the projected bulk valence bands at high temperature (400 K).
Such behaviour could be interpreted as a topological phase transition from topological phase at low temperature to normal one at high temperature.
The possible mechanisms to cause such transition, such as the thermal lattice distortion and electron-phonon coupling are examined.
Such new mechanism to undergo the topological phase transition might be useful to realize future spin-dependent thermoelectric devices.

\section*{Ackowledgements}
We thank T. Nakamura for his support during general experiments.
We also thank F. Deschamps for her support during the experiments on the CASSIOP\'EE beamline at synchrotron SOLEIL.
Part of the ARPES and LEED experiments were performed under the Nanotechnology Platform Program at IMS of the Ministry of Education, Culture, Sports, Science and Technology (MEXT),
This work was also supported by JSPS KAKENHI (Grants Nos. JP15H03676 and JP17K18757).

\begin{figure}[b]
\includegraphics[width=130mm]{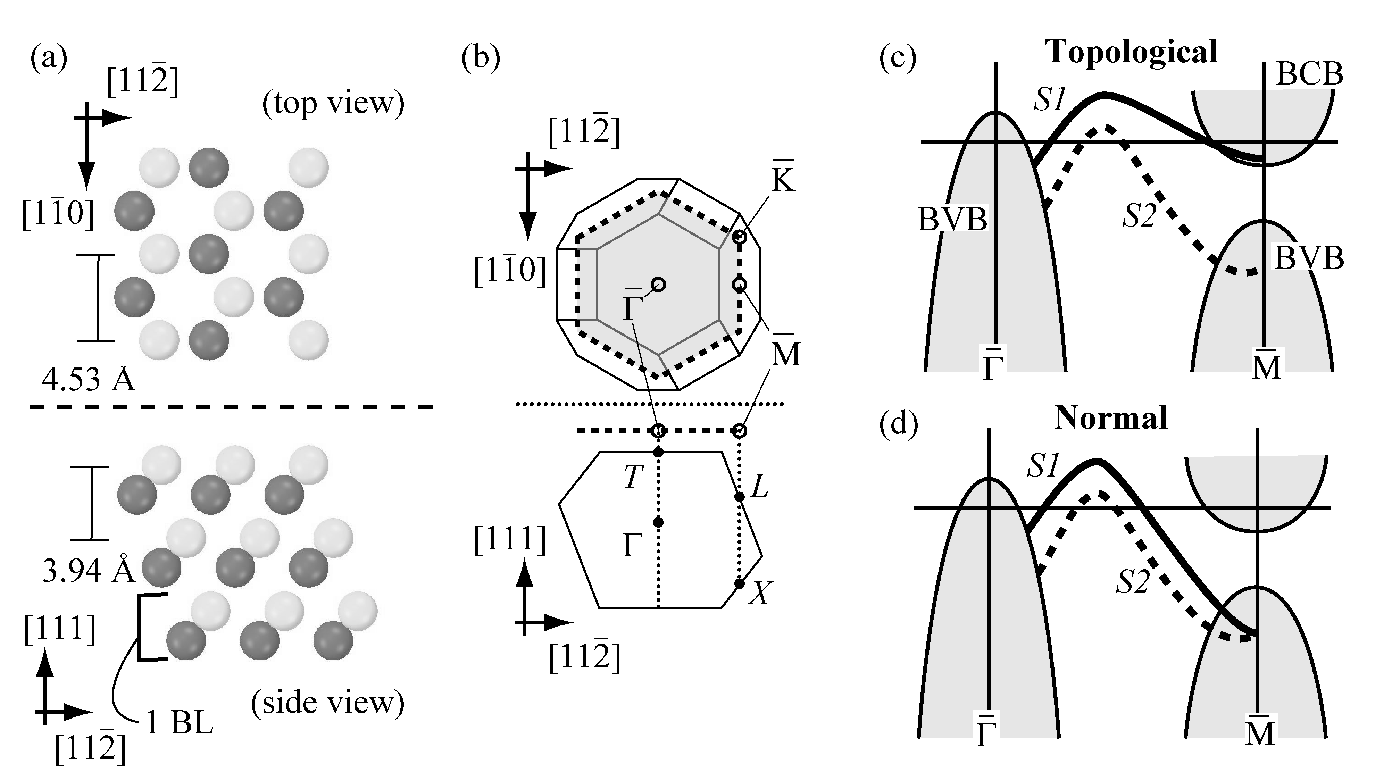}
\caption{\label{fig1}
(a) Atomic structure of single-crystal Bi.
(b) A schematic drawing of the three-dimensional Brillouin zone (solid) of the Bi single crystal and its projection to the (111) surface Brillouin zone (dashed).
(c) A schematic drawing of the surface bands and projected bulk bands on the (111) surface of the BiSb alloy  and Bi based on the topological case of bulk TO.
(d) The same as (c) but with the normal (non-topological) case of the bulk electronic structure.
}
\end{figure}

\begin{figure}
\includegraphics[width=80mm]{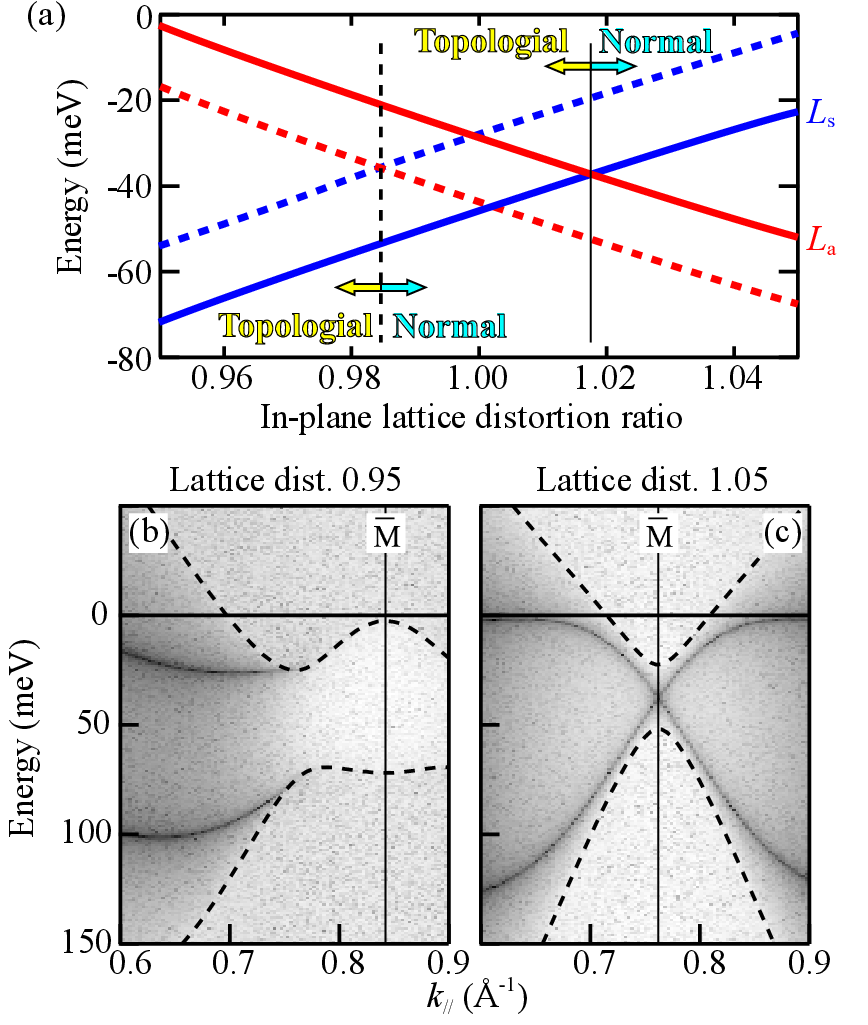}
\caption{\label{fig2}
(a) Band evolution at $L$ depending on the lattice distortion based on the empirical tight-binding model. Vertical lines indicate the position where the topological phase transition occurs.
The solid and dashed lines are based on the tight-binding parameter set which shows topological and normal TO of Bi without strain.
(b, c) Surface-state band dispersion calculated by the transfer-matrix method \cite{Teo08, Lee81}. The black area represents the surface-band dispersion and the dashed lines represents the edge of the projected bulk bands.
These figures are reproduced from the data shown in \cite{Ohtsubo16}.
}
\end{figure}

\begin{figure}[htbp]
\includegraphics[width=80mm]{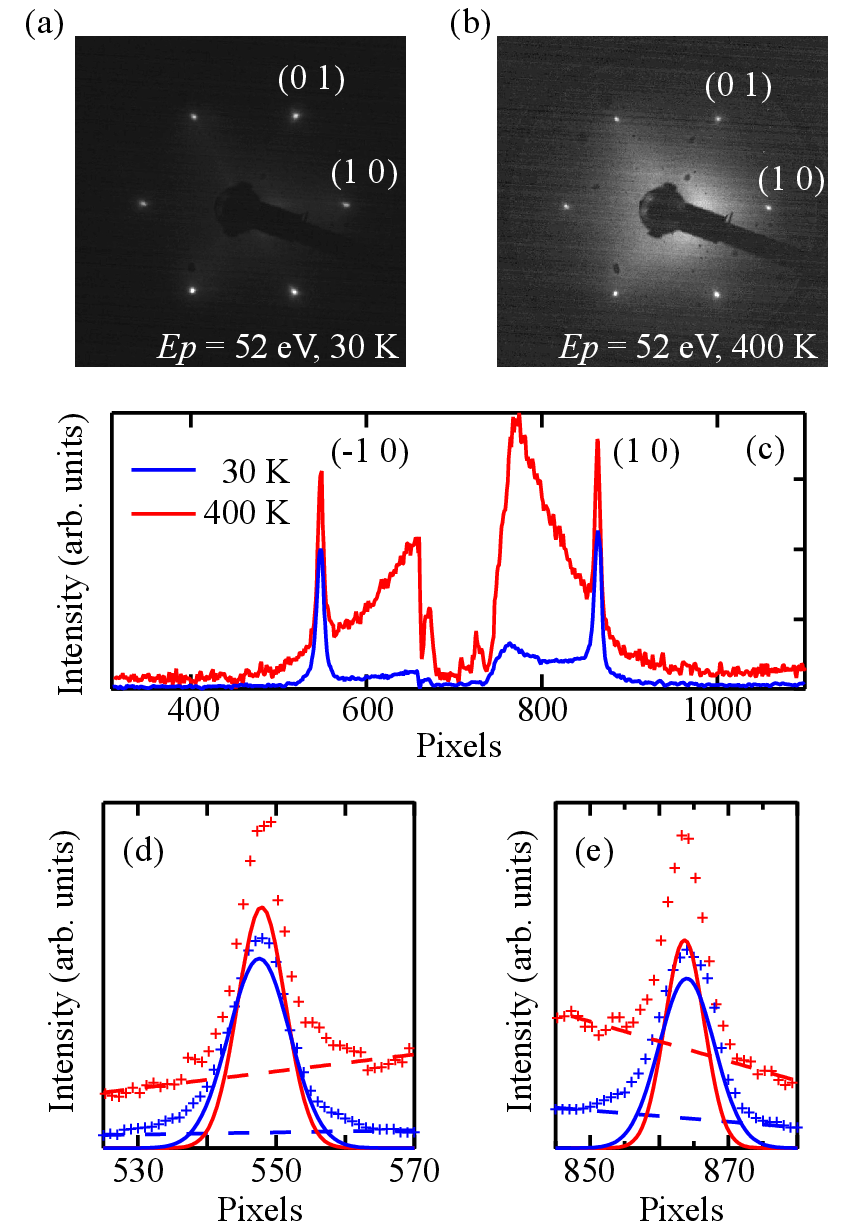}
\caption{\label{figs3}
(a, b) LEED patterns of the Bi(111) clean surface at (a) 30 K and (b) 400 K.
(c) LEED line profiles cutting (-1 0) and (1 0) (horizontal cut in (a) and (b)). The intensity vanishes at the centre of the profiles (670-730 pixels) because this area is shaded by an electron gun.
(d, e) The close-up image of the line profile shown in (c) around (d) (-1 0) and (e) (1 0).
The markers are the raw profile.
The dashed and solid lines are linear backgrounds and Gaussian peaks to fit the LEED line profile, respectively.
}
\end{figure}

\begin{figure}[htbp]
\includegraphics[width=80mm]{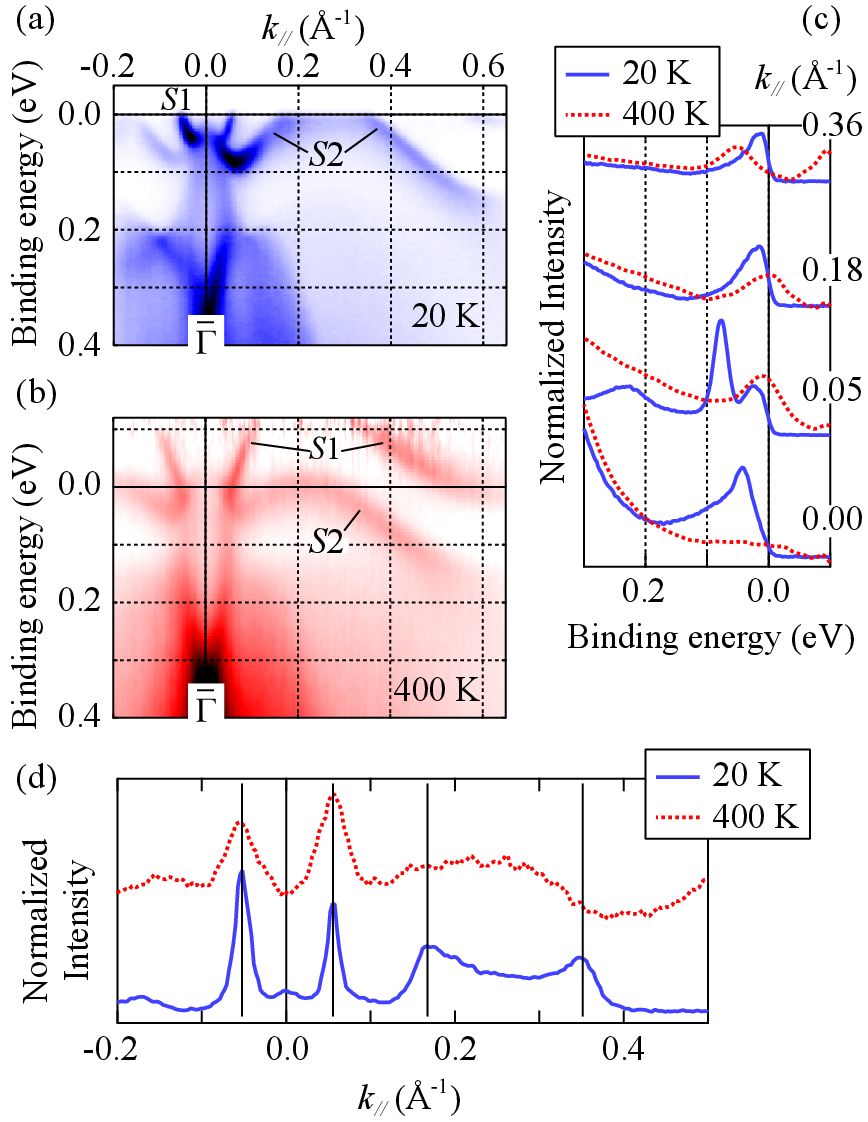}
\caption{\label{figs4} 
(a, b) ARPES intensity plots around $\bar{\Gamma}$ at (a) 20 K and (b) 400 K by linearly polarized photons at $h\nu$ = 30 eV.
The ARPES intensities in (b) are divided by the Fermi-Dirac distribution at 400 K convolved with the instrumental resolution.
(c) ARPES energy distribution curves (EDCs) at each in-plane wavevector $k_{\parallel}$ at both temperatures.
(d) ARPES momentum distribution curves at both temperatures obtained at the Fermi level. Vertical lines represents the $k_{\rm F}$ positions at 20 K.
}
\end{figure}

\begin{figure}
\includegraphics[width=80mm]{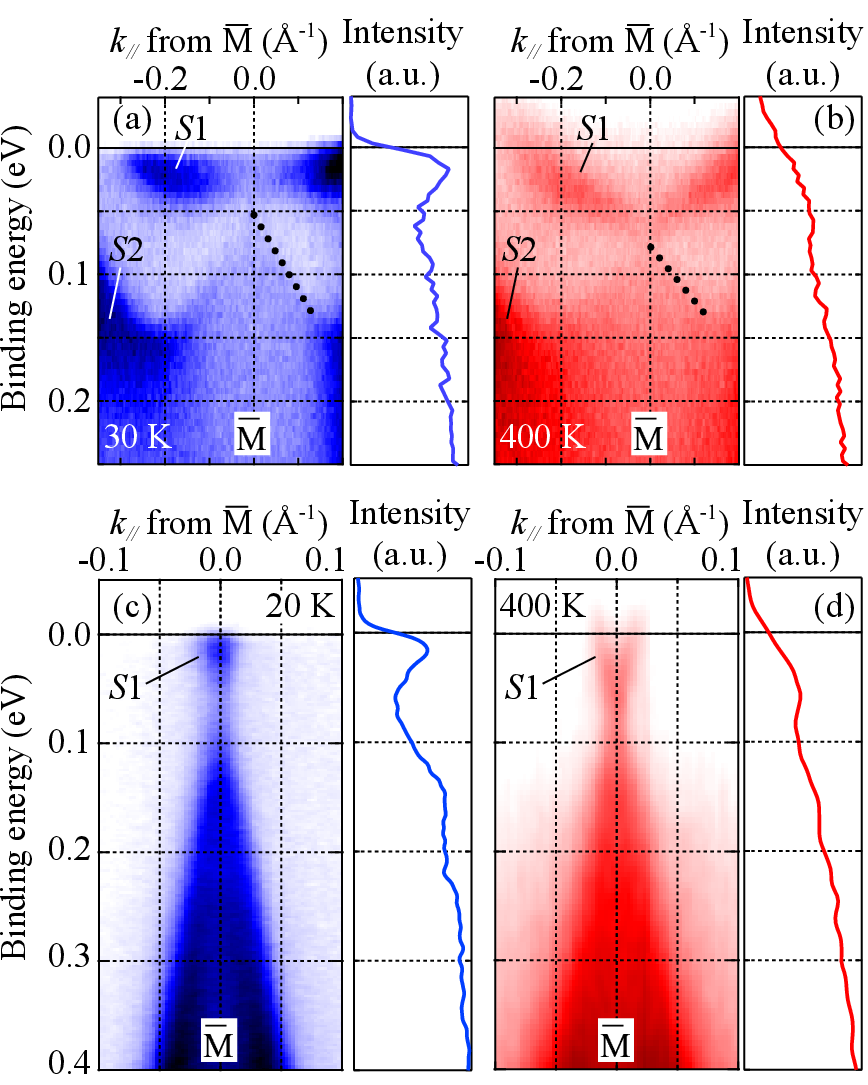}
\caption{\label{fig5}
(a, b) ARPES intensity plots along $\bar{\Gamma}$-$\bar{\rm M}$, around $\bar{\rm M}$ at (a) 30 K and (b) 400 K at $h\nu$ = 21.2 eV.
The spectra on the right side represent the EDCs at $\bar{M}$.
(c, d) The same as (a, b), respectively, but taken along $\bar{\rm M}$-$\bar{\rm K}$ at $h\nu$ = 30 eV at (c) 20 K and (d) 400 K.
}
\end{figure}

\begin{figure}
\includegraphics[width=160mm]{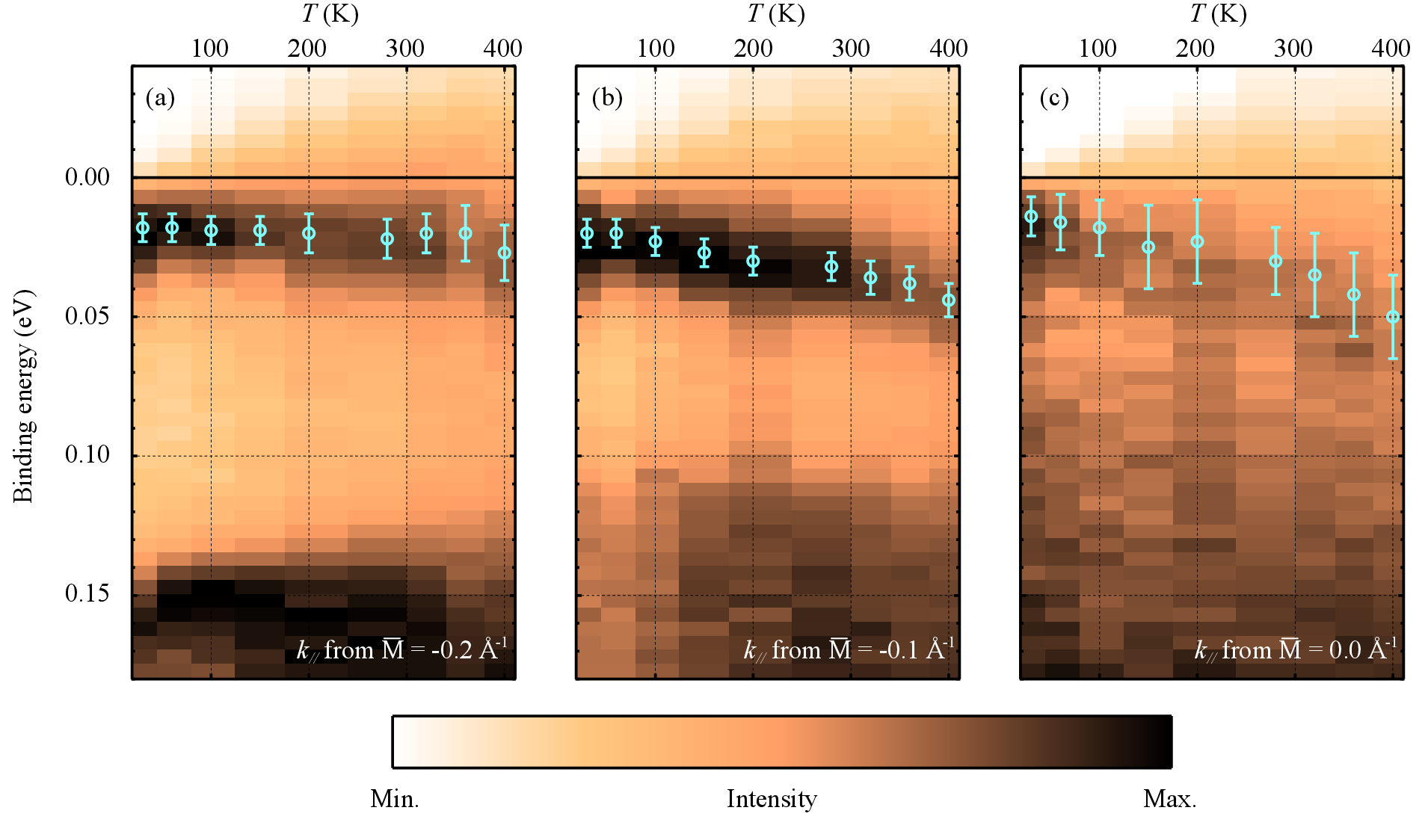}
\caption{\label{fig6}
ARPES intensity plots at the temperatures from 30 to 400 K at each $k_{\parallel}$ point. Circles indicate the peak positions of each EDC corresponding to the upper branch of the surface electronic state. The error bars are the possible range of the peak positions expected from the peak shape.
}
\end{figure}

\begin{table}[b]
\caption{\label{tab:two} Parity invariants ($\delta$) at each time-reversal-invariant momentum ($\Gamma , L, X, T$) and the $Z_2$ topological invariants ($\nu_0; \nu_1 \nu_2 \nu_3$) calculated according to the known theoretical models \cite{Fu07, Teo08}}
\begin{ruledtabular}
\begin{tabular}{cccccc}
 & $\delta$($\Gamma$) & $\delta$($L$) & $\delta$($X$) & $\delta$($T$) & ($\nu_0; \nu_1 \nu_2 \nu_3$) \\
\hline
Bi (normal) & $-$ 1 & $-$1 & $-$1 & $-$1 & (0;000) \\
BiSb (topological) & $-$1 & $+$1 & $-$1 & $-$1 & (1;111) \\
\end{tabular}
\end{ruledtabular}
\end{table}

\end{document}